# Battery-Assisted Operation of Hyperscale AI Data Centers under Connect-and-Manage Interconnection Practices

Xin Lu, *Senior Member*, *IEEE*, Jing Qiu, *Senior Member*, *IEEE*, Jiafeng Lin, *Student Member*, *IEEE*, Sihai An, *Student Member*, *IEEE*, Mingyang Sun, *Student Member*, *IEEE*, Junhua Zhao, *Senior Member*

***Abstract*—Emerging "connect-and-manage" practices allow new transmission-connected mega-loads to connect while enforcing time-varying admissible power exchange limits at the point of common coupling (PCC) in real time. Hyperscale artificial intelligence data centers (AIDCs), whose demand can reach hundreds of megawatts and whose internal computing–cooling dynamics evolve rapidly, can therefore face frequent conflicts between workload continuity requirements and externally imposed PCC envelopes. This paper proposes a battery-assisted operational framework in which on-site battery energy storage (BESS) serves as a physical buffering interface to reconcile fast internal dynamics with time-varying interconnection limits. A continuity-aware energy–computation model is developed to jointly capture checkpoint-constrained AI training workloads, information technology (IT) computing power–throughput characteristics, and IT–cooling thermal dynamics. A two-stage decision framework is then formulated, consisting of scenario-based day-ahead workload commitment and a real-time receding-horizon delivery assurance controller that enforces battery, thermal, and grid-interaction constraints. Case studies on the IEEE 39-bus system with Australian real data demonstrate that BESS substantially increases credible day-ahead workload commitment and improves real-time delivery robustness under transmission congestion. Sensitivity analyses further reveal a regime-dependent role transition of BESS—from feasibility-oriented continuity support when PCC limits are binding to economy-driven flexibility provision as transmission constraints are relaxed.**

## I. Introduction

In recent years, the rapid growth of large-scale artificial intelligence training workloads has turned hyperscale artificial intelligence data centers (AIDCs) [1] into some of the highest-power and fastest-growing grid-connected loads in modern power systems [2]. A single AIDC can demand several hundred megawatts of electricity, posing unprecedented challenges to transmission network capacity and operational security [3]. However, the planning and construction timescales of transmission infrastructure are often significantly longer than the deployment cycles of AIDCs themselves [4].

Under conventional large-load interconnection practices, power systems typically adopt a "review-first, connect-later" paradigm, where network feasibility is resolved ex ante through detailed studies and reinforcements. For AIDCs with short construction cycles and rapid capacity expansion, this paradigm can substantially delay commissioning and struggle to keep pace with the accelerating demand for AI computing. As a result, recent industry and regulatory discussions have begun to explore interconnection approaches that rely more heavily on operational-stage management. For example, on January 16, 2026, PJM Interconnection proposed a "connect-and-manage" framework under which new large loads may be curtailed during periods of system stress to preserve transmission reliability [5]. This shift moves feasibility considerations from the connection stage to real-time operation, raising fundamental questions about how AIDC power behavior should be coordinated with grid constraints.

Existing studies on large-scale load coordination have primarily focused on demand response and load flexibility. One line of work treats data centers as controllable loads [6, 7] and leverages electricity price signals [8] or incentive mechanisms [9] to guide power adjustments for peak shaving or cost reduction. Another line of research emphasizes load-side power smoothing and ramp-rate control, aiming to mitigate the impact of rapid load variations on grid operational security [10]. While these approaches highlight the potential of data center loads to participate in system-level regulation, they typically rely on implicit assumptions that loads can be switched on and off at arbitrary times and that short-duration interruptions have negligible effects on internal operations.

These assumptions do not hold for AIDCs dominated by large-scale AI training workloads. AI training is inherently stateful and can only be safely interrupted at discrete checkpoint moments [11]. Moreover, information technology (IT) computing loads are tightly coupled with cooling systems through heat generation and dissipation, giving rise to fast internal dynamics involving multiple interacting subsystems [12]. As a result, AIDC behavior cannot be adequately represented as a single, freely adjustable power variable.

From the grid perspective, interconnection constraints are typically enforced as power exchange limits at the point of common coupling (PCC), reflecting transmission thermal ratings and real-time system conditions [13]. These limits are inherently time-varying and governed by network-wide power flow patterns rather than the internal dynamics of an individual load. Consequently, the common assumption that load-side flexibility can be directly mapped to feasible grid-side regulation becomes problematic for ultra-large loads with fast internal dynamics such as AIDCs [14, 15].

Against this backdrop, a critical question arises: how can operational continuity of an AIDC be ensured when feasible PCC power varies over time and may temporarily tighten [16]? Reliance on direct load curtailment or shutdown risks frequent computational interruptions and undermines the ability of AIDCs to continuously deliver AI workloads under constrained conditions.

In this context, physical mechanisms that provide time-scale decoupling and power buffering between the grid and the load become essential. Battery energy storage systems (BESSs) [17], with fast power response and bidirectional energy regulation capabilities, have been widely used to smooth power fluctuations and alleviate network constraints [18]. When interconnection limits are time-varying and occasionally binding, BESSs offer the potential to reconcile AIDC power demand with transmission feasibility without directly disrupting internal computing operation.

Beyond operational feasibility, hyperscale AIDCs are service-oriented infrastructures that must plan and commit computing workloads in advance [19, 20]. Such commitments are typically made based on anticipated power availability, while actual interconnection limits may evolve during real-time operation [21]. This disconnect exposes AIDCs to delivery risks when real-time power availability deviates from expectations, highlighting the need for coordinated day-ahead planning and real-time operational assurance.

Existing studies on data center operation remain inadequate for hyperscale AIDCs dominated by large-scale AI training under power-constrained conditions. In particular, prevailing models often fail to jointly capture the stateful, checkpoint-constrained nature of AI training and the tightly coupled IT–cooling dynamics that are essential for representing operational continuity. Furthermore, under time-varying transmission-level interconnection limits, the lack of an explicit operational interface to align fast internal dynamics with externally imposed power constraints limits the applicability of existing approaches. More generally, within “connect-and-manage” interconnection practices, the joint problem of advance workload commitment and real-time operational feasibility under uncertain power availability has not been systematically addressed:

1. To accommodate the emerging “connect-and-manage” interconnection paradigm for future hyperscale AIDCs, an energy–computation modelling approach is presented, which jointly characterizes stateful, checkpoint-constrained AI training workloads and tightly coupled IT–cooling thermal dynamics, providing a coherent representation of AIDC operational continuity under power-constrained conditions.
2. A novel grid-aware operational framework is proposed for AIDCs operating under the “connect-and-manage” regime. On-site BESSs are positioned as a physical buffering interface, enabling time-scale decoupling between fast internal computing–cooling dynamics and time-varying transmission-level interconnection constraints.
3. A two-stage decision logic is developed to coordinate workload commitment with real-time grid feasibility. The logic consists of a scenario-based day-ahead workload commitment mechanism and a real-time receding-horizon delivery assurance controller, ensuring that AI workload commitments remain credible under uncertain transmission congestion while maintaining strict compliance with grid operational constraints.

The remainder of this paper is organized as follows. Section II introduces the proposed battery-assisted operational framework for hyperscale AIDCs. Section III presents the energy–computation model. Section IV formulates the two-stage decision logic for day-ahead workload commitment and real-time delivery assurance under connect-and-manage practices. Section V provides case studies and sensitivity analyses, and Section VI concludes the paper.

## II. Battery-Assisted Operational Framework for AIDCs Under Connect and Manage

We consider a hyperscale AIDC as a new class of grid-connected mega-load that is both power-intensive and operationally flexible, yet increasingly constrained by transmission-level limits under emerging “connect-and-manage” interconnection practices. Unlike conventional data centers dominated by latency-sensitive services, modern AIDCs are largely driven by batch-oriented AI training: the computation can be throttled in software, shifted in time, and even paused when needed. However, this flexibility is not “free.” Large-scale training is stateful and must respect continuity requirements, for example, training progress can only be safely interrupted at checkpoint-compatible moments. In parallel, the AIDC’s electrical demand is tightly coupled with its thermal management: cooling effort must respond to heat generation and ambient conditions, which introduces additional short-timescale variability. Under “connect-and-manage” operation, transmission congestion is enforced by the transmission system operator (TSO) as time-varying admissible power exchange limits at the PCC, which may tighten or relax dynamically during system operation. If the grid-facing exchange were forced to directly track these internal fluctuations, the AIDC would frequently conflict with interconnection limits, leading to unreliable workload delivery on the AIDC side and increased operational risk on the grid side. In other words, the grid “sees” volatility, while the AIDC “feels” constraints, and the two objectives can become incompatible without an appropriate interface.

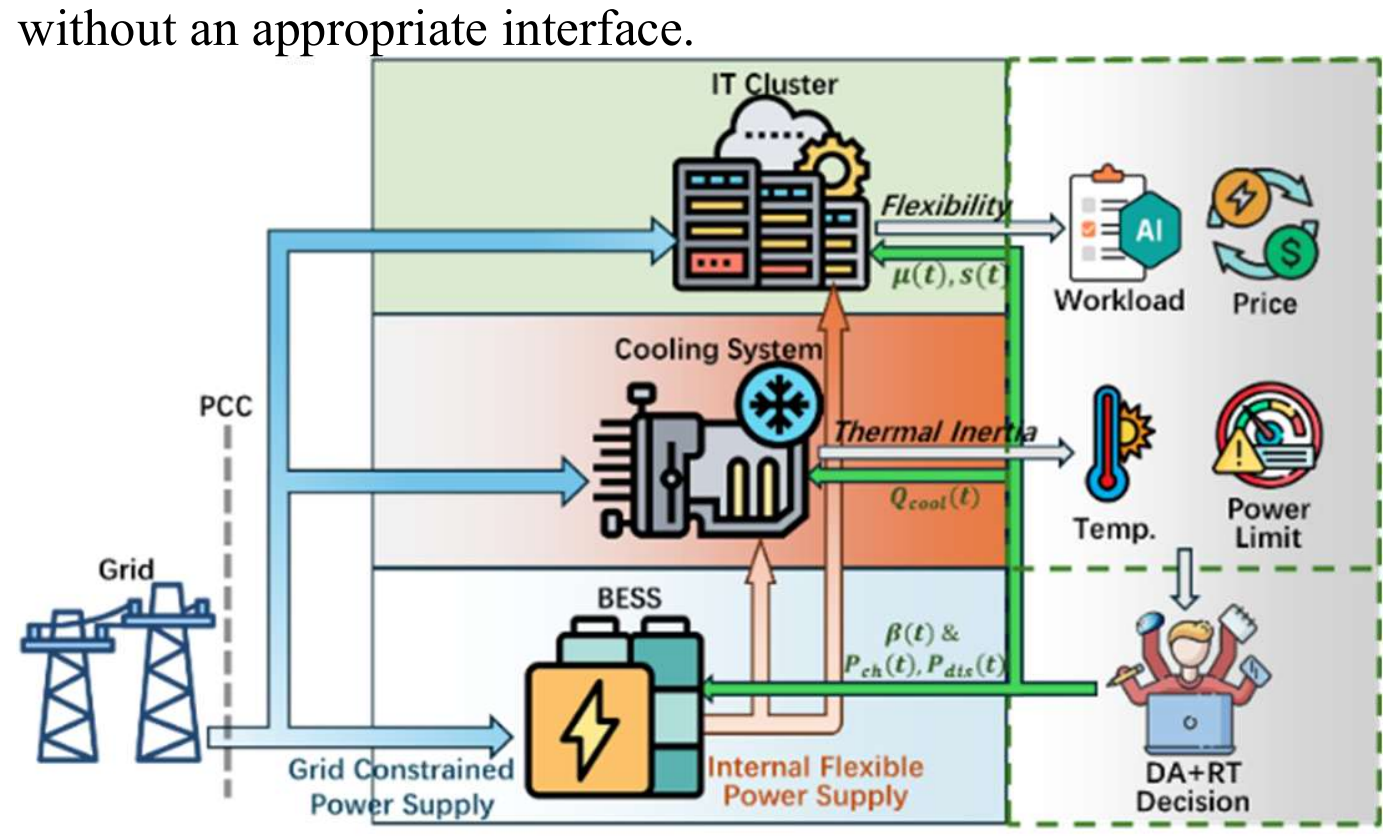

Fig. 1 Battery-Assisted AIDC Operational Framework

To resolve this mismatch, we adopt a battery-assisted operational framework (Fig. 1) that deliberately positions the on-site BESS as the central physical interface between the power grid and the AIDC’s internal subsystems. The key idea is simple but powerful: under connect-and-manage interconnection, the grid-facing power exchange is dictated by operator-imposed limits rather than by the AIDC’s internal dynamics. As a general operational principle, the grid should not be asked to follow fast internal variations. Instead, the BESS absorbs short-term mismatches between what the AIDC needs internally and what the grid can safely accommodate externally. When grid capacity is temporarily tight due to binding PCC limits, the BESS sustains internal operation by providing fast electrical buffering. When grid capacity is ample, or when prices are favorable, the BESS can recharge or opportunistically support economically beneficial actions, while still maintaining strict compliance with interconnection requirements. In this role, the BESS is treated as the dispatchable storage asset coordinating energy exchange and internal demand; embedded UPS units within IT hardware are not relied upon for system-level operational scheduling.

This physical decoupling enables coordinated flexibility across three tightly coupled layers. At the computing layer, software-defined power limits allow throughput to be modulated to follow grid conditions and market signals, while continuity requirements restrict when shutdowns are permissible—making battery buffering critical during “can’t-stop” periods. At the thermal layer, the decoupled operation allows the system to exploit thermal inertia, shifting cooling effort in time relative to heat generation without violating temperature constraints, effectively reallocating electrical power between IT and cooling during constrained intervals. the grid interface, the AIDC behaves as a predictable, controllable resource whose external power exchange respects TSO-enforced, time-varying PCC limits and ramping requirements, even when internal subsystems are fluctuating.

By unifying computational flexibility, checkpoint-constrained continuity, thermal inertia, and electrical storage through a single physical interface, the proposed framework transforms the AIDC from a volatile mega-load into a grid-compatible and economically responsive prosumer under connect-and-manage practices, enabling reliable workload delivery while honoring transmission-level interconnection constraints.

## III. Energy-Computation Modelling of AIDC Dynamics

### A. Hyperscale AIDC Power Consumption

The proposed hyperscale AIDC is modeled as a controllable high-power load connected to a designated transmission bus, denoted as the location bus (loc). The net active power exchange at the PCC between the transmission grid and the AIDC at time $t$, denoted as $P_{AIDC}^{exc}(t)$ (positive for import, negative for export), is determined by the aggregate power demand of its internal subsystems:

$$P_{AIDC}^{exc}(t) = \frac{P_{IT}(t)}{\eta_{IPCS}} + P_{cool}(t) + P_{bess}(t) \tag{1}$$

where $P_{IT}(t)$ denotes the dynamic IT power consumption, $\eta_{IPCS}$ represents the aggregate efficiency of the internal power conditioning system, $P_{cool}(t)$ denotes the electrical power required by the cooling system, and $P_{bess}(t)$denotes the charging (positive) or discharging (negative) power of the on-site BESS.

### B. DCAI Computing Model

#### 1) Computing Cluster

The computational core consists of $N_{server}$ homogeneous GPU servers processing delay-tolerant, batch-oriented AI training workloads. The IT cluster is controlled through software-defined power limits, which regulate the effective computational throughput and the electrical load. The operational state is described by a binary activation indicator $\mu(t) \in \{0,1\}$, representing whether the computing cluster is active, and a normalized throughput variable $s(t) \in [0,1]$. The coupling between activate state and throughput is enforced by:

$$s_{min}\mu(t) \leq s(t) \leq \mu(t) \tag{2}$$

where $s_{min}$ denotes the minimum throughput level when the computing cluster is active. Therefore, (2) ensures that the throughput is zero when the cluster is inactive ($\mu(t) = 0$), and bounded between $s_{min}$ and 1 when the cluster is active($\mu(t) = 1$). The instantaneous workload processing rate is proportional to the normalized throughput when the cluster is active, and is given by:

$$W(t) = N_{server} r_{peak} s(t)\mu(t) \tag{3}$$

where $r_{peak}$ denotes the peak throughput of a single server.

#### 2) Real-Power Consumption

Although computational throughput is regulated via software-defined power capping, empirical measurements [22] indicate that the electrical power drawn by modern AI accelerators exhibits a strictly convex dependence on throughput. Accordingly, the aggregate IT power consumption is modeled as:

$$P_{IT}(t) = N_{server}\mu(t)(\alpha_0 + \alpha_1 s(t) + \alpha_2 s(t)^2) \tag{4}$$

where $\alpha_0, \alpha_1, \alpha_2$ are empirical coefficients obtained by fitting the measured power–throughput operating points over the admissible throughput range. The quadratic form serves as a local surrogate model and is not extrapolated beyond the calibrated region.

#### 3) Checkpoint Constraints

Large-scale AI training is stateful and requires periodic checkpointing to preserve computational progress. We introduce an exogenous binary parameter $\delta_{ckpt}(t)$, which indicates whether a time slot permits shutdown. We regulate the evolution of the activate state $\mu(t)$ by constraining its inter-temporal transitions as follows:

$$\mu(t) \geq \mu(t-1) - \delta_{ckpt}(t) \tag{5}$$

Constraint (5) enforces checkpoint-aware operational continuity. When $\delta_{ckpt}(t) = 0$, the activate state is required to satisfy $\mu(t) \geq \mu(t-1)$, thereby prohibiting transitions from an active state $\mu(t-1) = 1$ to a shutdown state $\mu(t) = 0$. Conversely, when $\delta_{ckpt}(t) = 1$, the constraint becomes non-binding and allows controlled shutdowns at admissible checkpointing windows. Notably, this constraint only restricts shutdown actions; start-up is assumed feasible on the 15-min timescale via fast job dispatching and cluster power capping; the dominant constraint is checkpoint-dependent shutdown to avoid progress loss.

### C. Thermal Dynamics and Cooling Power Model

#### 1) Thermal State Dynamics

The thermal state of the data hall is regulated through a continuous control variable, $Q_{cool}(t)$, which represents the thermal power extracted by the cooling system. Let $T_{in}(t)$ denote the indoor temperature at time slot $t$. A first-order equivalent thermal circuit (RC) model is employed to describe the temperature evolution from time slot from $t$ to $t+1$:

$$\mathcal{T}_{in}(t+1) = \mathcal{T}_{in}(t) + \frac{\Delta t}{C_{th}}\left(P_{IT}(t) - \frac{\mathcal{T}_{in}(t) - \mathcal{T}_{amb}(t)}{R_{th}} - Q_{cool}(t)\right) \tag{6}$$

where $C_{\text{th}}$ denotes the aggregate thermal capacitance of the facility, including servers, air, and infrastructure, and $R_{\text{th}}$ represents the thermal resistance between the indoor environment and the ambient temperature $T_{\text{amb}}(t)$. $\Delta t$ denotes the control time step. The cooling extraction power $Q_{cool}(t)$, is limited by the rated cooling capacity of the facility, denoted by $Q_{cool}^{max}$.

#### 2) Electrical Cooling Power

The electrical power consumption of the cooling system, denoted by $P_{cool}(t)$, represents the electrical input required to remove the thermal load from the data center:

$$P_{cool}(t) = \text{EIR}(t) Q_{cool}(t) \tag{7}$$

$$\text{EIR}(t) = \text{EIR}_{nom}\phi(t) \tag{8}$$

where $\text{EIR}(t)$ and $\text{EIR}_{nom}$ denote the temperature-dependent energy input ratio and the nominal value, and $\phi(t)$is a normalized temperature-dependent correction factor [23]:

$$\phi(t) = -0.000006\, T_{amb}^{F}(t)^2 + 0.004941 T_{amb}^{F}(t) + 0.58462 \tag{9}$$

where $T_{amb}^{F}(t)$is the outdoor air temperature expressed in degrees Fahrenheit (°F), obtained from the ambient temperature in degrees Celsius via

$$T_{amb}^{F}(t) = 1.8\, T_{amb}(t) + 32 \tag{10}$$

#### 3) Thermal Comfort Constraints

To ensure safe and reliable operation, the indoor temperature is constrained within an admissible range:

$$\mathcal{T}_{min} \leq \mathcal{T}_{in}(t) \leq \mathcal{T}_{max} \tag{11}$$

## D. Battery Energy Storage System Model

### 1) Power and Energy Dynamics

The storage dispatch is described using a binary operating mode indicator $\beta(t) \in \{0,1\}$, together with two non-negative continuous variables $P_{ch}(t)$and $P_{dis}(t)$, representing the charging and discharging power, respectively. The BESS power balance is given by:

$$P_{bess}(t) = P_{ch}(t) - P_{dis}(t) \quad (12)$$

Let $E_{bess}(t)$ denote the State of Charge (SoC) energy level at $t$. The dynamics are modeled as a first-order difference equation:

$$E_{bess}(t+1) = E_{bess}(t) + \left(P_{ch}(t)\eta_{ch} - \frac{P_{dis}(t)}{\eta_{dis}}\right)\Delta t \quad (13)$$

where $\eta_{ch}$ and $\eta_{dis}$ represent the charging and discharging efficiencies, respectively.

### 2) Power and Energy Constraints

The operation is bounded by the maximum rated power $P_{bess}^{max}$:

$$0 \leq P_{ch}(t) \leq \beta(t) \cdot P_{bess}^{max} \quad (14)$$

$$0 \leq P_{dis}(t) \leq \left(1 - \beta(t)\right) \cdot P_{bess}^{max} \quad (15)$$

where binary mode variable $\beta(t)$ enforces mutually exclusive charging/discharging. To ensure operational safety and limit excessive degradation, the state of charge is restricted to a prescribed range:

$$E_{min} \leq E_{bess}(t) \leq E_{max} \quad (16)$$

Finally, to guarantee cyclic consistency over the scheduling horizon, the terminal state of charge is constrained to match the initial condition as follows. This cyclic constraint avoids end-of-horizon energy depletion and ensures fair comparison across scenarios.

$$E_{bess}(T+1) = E_{bess}(1) \quad (17)$$

## E. Transmission Network and Grid Interface

To characterize the physical origin of grid-interaction constraints at the PCC, we first describe a representative transmission network model typically used by the TSO. A linearized DC power flow representation is considered for illustration. Let $\mathcal{N}$ denote the set of buses and $\mathcal{L}$ the set of transmission lines. For every bus $k \in \mathcal{N}$, the nodal active power balance can be written as:

$$P_{Gen,k}(t) - P_{Load,k}(t) - \delta_{k,loc} P_{AIDC}^{exc}(t) = \sum_{j \in \Omega_k} P_{line,kj}(t) \quad (18)$$

where $P_{Gen,k}(t)$ and $P_{Load,k}(t)$ represent the background generation and load at bus $k$. The binary indicator $\delta_{k,loc}$ is 1 if the AIDC is located at bus $k$, and $\Omega_k$ denotes the set of buses connected to bus $k$. $P_{line,kj}(t)$ represents active power flow on the transmission line $(k,j) \in \mathcal{L}$. The DC power flow relations and line thermal constraints underlying (18), which are standard and used by the TSO to assess network congestion, are not solved by the AIDC and are therefore provided in Appendix A for illustrative purposes. Based on a PTDF-based sensitivity analysis of line thermal limits, the impact of transmission network constraints on the AIDC is internalized in the form of time-varying admissible power exchange limits at the PCC:

$$\underline{P_{grid}}(t) \leq P_{AIDC}^{exc}(t) \leq \overline{P_{grid}}(t) \quad (19)$$

From the AIDC operator's perspective, the admissible power exchange limits $\underline{P_{grid}}(t)$ and $\overline{P_{grid}}(t)$ at the PCC are treated as exogenously imposed constraints, regardless of the underlying physical or operational causes. In addition to these time-varying amplitude limits, operational practice also imposes constraints on how fast the power exchange can change over time. Accordingly, the TSO-imposed ramp-rate requirement is represented by a ramp limit $R_{grid}$, defined as the maximum allowable change in the PCC net power exchange between two consecutive 15-min time slots (MW per time step).

# IV. Two-Stage Decision Logic for Workload Commitment and Delivery Assurance

## A. Two-Stage Decision Logic Under Connect-and-Manage

Under "connect-and-manage" practices, large grid-connected loads are permitted to connect subject to operationally enforced power exchange limits at PCC. These limits are communicated by the system operator as binding constraints and are updated in real time as system conditions evolve. As a result, the AIDC operator faces an asymmetric information structure across time scales: in real time, the operator receives the realized admissible PCC power exchange limits, which must be satisfied as hard constraints; in the day-ahead stage, the operator can only rely on historical observations and publicly observable signals—such as electricity prices, system demand, and ambient temperature—to infer how the admissible PCC power exchange limits are likely to evolve. To separate commitment credibility from operational execution under this information asymmetry, we adopt a two-stage decision logic consisting of:

**Stage I (Day-Ahead, DA):** determine how much AI workload can be credibly committed in advance based on publicly observable signals and historical operator-imposed limits;

**Stage II (Real-Time, RT):** ensure delivery of the committed workload under realized system operator constraints and disturbances via receding-horizon control.

Crucially, the DA stage does not prescribe executable power or battery trajectories. Instead, it determines a single scalar commitment $W^{DA*}$, representing the amount of AI workload promised for completion within the delivery horizon. All detailed operational decisions, including computing throttling, cooling control, and BESS dispatch, are deferred to real time, where the system operator communicates binding PCC power exchange limits and prices. These variables are denoted as $x(t) = \{\mu(t), s(t), Q_{\text{cool}}(t), P_{\text{ch}}(t), P_{\text{dis}}(t), \beta(t)\}$.

## B. Day-Ahead Scenario-Based Workload Commitment

As discussed in Section IV.A, an AIDC operator must rely on historical observations and publicly observable system signals to infer how these limits are likely to evolve, solely for the purpose of credible workload commitment. This section develops a data-driven power exchange limit prediction model and uses the inferred limits to determine a high-confidence day-ahead workload commitment.

Using the historical dataset, we learn the relationship between publicly observable system stress indicators $z(t)$ and the admissible PCC power exchange limits $\underline{P_{grid}}(t)$ and $\overline{P_{grid}}(t)$ imposed by the TSO. The feature vector $z(t) = [\pi(t), D(t), T_{\text{amb}}(t), h(t)]$ includes electricity price, demand, ambient temperature, and calendar indicators. Rather than predicting single-point values of the admissible PCC limits, we adopt a conditional generative modelling approach to explicitly capture the uncertainty and temporal correlation inherent in operator-imposed power limits. The generative model is trained to learn the conditional joint distribution and to generate a set of admissible power limit trajectories $\left\{\underline{P_{grid}^{\omega}}(t), \overline{P_{grid}^{\omega}}(t)\right\}_{t=1}^{T}, \omega \in \Psi_{raw}$, that are statistically consistent with historical operator behavior under similar system conditions. To control the level of conservatism, a target coverage level $1-\alpha$ is specified. From the raw ensemble $\Psi_{raw}$, only the central $1-\alpha$fraction of power limit trajectories is retained, yielding a reduced scenario set $\Psi \subset \Psi_{raw}$, which excludes extreme tail realizations while preserving representative variability in admissible PCC limits. This retained scenario set $\Psi$ is used exclusively for day-ahead workload commitment.

The objective of the DA stage is to determine a single scalar workload commitment that can be delivered with high confidence across all retained admissible power limit scenarios. To ensure economic and operational credibility, the commitment decision is anchored at the energy-efficient throughput point $s^*$, defined as the operating point that maximizes the amount of workload processed per unit electrical power. Deviations from this efficiency-optimal operating regime are penalized but not prohibited, allowing flexibility while discouraging commitments that rely on inefficient or extreme operation. Let $W^{DA}$ denote the committed workload, and let $x^{\omega}(t)=\{\mu^{\omega}(t), s^{\omega}(t), Q^{\omega}_{\mathrm{cool}}(t), P^{\omega}_{\mathrm{ch}}(t), P^{\omega}_{\mathrm{dis}}(t), \beta^{\omega}(t)\}$ collect the scenario-dependent internal operating decisions. The day-ahead workload commitment problem is formulated as:

$$\begin{aligned}\max_{W^{DA},\{x^{\omega}(t)\}} \; & W^{DA} - \lambda \textstyle\sum_{\omega\in\Psi}\sum_{t=1}^{T}|s^{\omega}(t)-s^*| \\ s.t.\; & \textstyle\sum_{t=1}^{T} N_{server} r_{peak} s^{\omega}(t)\mu^{\omega}(t)\,\Delta t \ge W^{DA}, \forall\omega\in\Psi, \\ & \underline{P^{\omega}_{grid}}(t) \le P^{exc,\omega}_{AIDC}(t) \le \overline{P^{\omega}_{grid}}(t), \forall t, \forall \omega\in\Psi, \\ & and\ constraints\ in\ Section\ III. \end{aligned} \tag{20}$$

where $P^{exc,\omega}_{AIDC}(t)$denotes the PCC power exchange under scenario $\omega$. The first constraint enforces scenario-wise deliverability: the same committed workload $W^{DA}$ must be achievable under every admissible power limit scenario retained after confidence filtering. As a result, the optimal solution corresponds to the largest workload level that is simultaneously deliverable across all considered scenarios, effectively representing a high-confidence commitment consistent with the inferred admissible grid interaction limits. Solving the above problem yields the day-ahead committed workload $W^{DA*}$, which is subsequently passed to the real-time stage. No operational trajectories determined in the day-ahead stage are binding in real time; all detailed decisions are deferred to real-time receding-horizon control under realized system operator constraints.

### C. Real-Time Receding-Horizon Optimization-based Delivery Assurance

At each 15-minute control interval, the AIDC obtains the realized electricity price $\pi_{RT}(t)$ together with the system operator–provided constraints $\underline{P^{exc,RT}_{grid}}(t)$ and $\overline{P^{exc,RT}_{grid}}(t)$, as well as the ramp-rate requirement $R_{grid}$. A receding-horizon optimization with horizon length $H$ is solved to ensure delivery of the previously committed workload $W^*_{DA}$ while minimizing realized operating cost.

To embed DA workload commitment into real-time decision-making, we define the remaining committed workload state $R(t)$:

$$R(t+1) = R(t) - W(t)\Delta t - S_{shed}(t) \tag{21}$$

$$R(1) = W^*_{DA},\; R(T+1) = 0 \tag{22}$$

where $S_{shed}(t) \ge 0$ is a nonnegative under-delivery slack activated only when real-time infeasibility prevents meeting the commitment. At control step $\tau$, the real-time problem minimizes:

$$\min J_{RT}(\tau) = \sum_{t=\tau}^{\tau+H-1}\begin{pmatrix}\pi_{RT}(t)P^{exc,RT}_{AIDC}(t)\Delta t + C_{de}\left(P^{RT}_{ch}(t)+P^{RT}_{dis}(t)\right)\Delta t \\ +M_{RT}S_{shed}(t)\end{pmatrix} \tag{23}$$

where $M_{RT}$ is a sufficiently large penalty coefficient applied to real-time workload under-delivery $S_{shed}(\tau)$. At each control step, only the first control action is implemented, after which the system states are updated and the optimization horizon is shifted forward. This receding-horizon execution enables the AIDC to adapt to real-time price volatility, PCC power exchange limits tightening, and thermal disturbances, while ensuring delivery of the workload committed in the DA stage. The overall battery-assisted two-stage grid-aware operation of the hyperscale AIDC is summarized in Algorithm 1.

**Algorithm 1:** *Battery-Assisted Two-Stage Grid-Aware Operation of a Hyperscale AIDC*

**Input:** Scheduling horizon $t=1,\dots,T$; control step $\Delta t$; Horizon $H$; AIDC, BESS, and thermal parameters; checkpoint availability $\delta_{ckpt}(t)$, historical dataset $\left\{z(t), \underline{P_{grid}}(t), \overline{P_{grid}}(t)\right\}$.

**Output:** Day-ahead committed workload $W^*_{DA}$; real-time executed trajectories and delivery record.

**Stage I:** *DA Scenario-Based Workload Commitment*

1. Train a conditional generative model that maps publicly observable features $z(t)$ to admissible PCC power exchange limit trajectories.
2. Generate $N_{raw}$ admissible power limit trajectories from the generative model, obtaining $\Psi_{raw}$. Discard tail scenarios using empirical quantile filtering at level $\alpha$, and retain a reduced scenario set $\Psi \subset \Psi_{raw}$.
3. Solve the DA workload commitment problem:
   **Maximize** $\boldsymbol{W^{DA} - \lambda \sum_{\omega,t}|s^{\omega}(t) - s^*|}$
   Subject to for every scenario $\omega \in \Psi$, there exists a feasible operating policy $\boldsymbol{x^{\omega}}$ satisfying:
   (a) AIDC power balance and subsystem models (1)–(4) and (6)-(17);
   (b) Checkpoint-aware operational continuity constraints (5);
   (c) Transmission-network impacts internalized via grid power limits and PCC ramp-rate requirements (19);
   (d) Scenario-wise workload deliverability constraint.
4. Set $W^*_{DA} \leftarrow W^{DA}$ and pass to the real-time stage.

**Stage II:** *RT Receding-Horizon Delivery Assurance*

5. Initialize remaining committed workload $R(1) \leftarrow W^*_{DA}$ and physical states ($\mathcal{T}_{in}(1)$, $E_{bess}(1)$, $\mu(0)$).
6. for $\tau = 1,\dots,T$ do
7. Obtain realized electricity price and the TSO provided grid power limits.
8. Solve a receding-horizon real-time optimization over $t = \tau,\ ,\tau+H-1$:
   **Minimize** $\boldsymbol{J_{RT}(\tau)}$ **defined in (23)**
   Subject to:
   a) AIDC power balance and subsystem models (1)–(4) and (6)-(17);
   b) Checkpoint-aware operational continuity constraints (5);
   c) Transmission-network impacts internalized via grid power limits and PCC ramp-rate requirements (19);
   d) Remaining-workload dynamics and terminal condition (21)–(22).
9. Implement only the first-step control actions at τ.
10. Update thermal state, battery state of charge using (6), (13), and the workload update equation, respectively.
11. end for

## V. Case Study

### A. Data Description and Study Setup

We conduct case studies with a scheduling horizon of $T = 96$ time slots, each representing a 15-minute interval ($\Delta t = 0.25$ h).

The external grid is represented by the IEEE 39-bus transmission test system (MATPOWER case39 via PYPOWER). The hyperscale AIDC is connected to bus $loc = 20$ . Transmission congestion impacts are internalized through a time-varying admissible power exchange power limits at PCC, provided by TSO. The power limits are computed using a DC power flow approximation based on background system demand and line thermal limits. The maximum import capacity at the PCC is limited to 1000 MW, while power export is allowed within the power limits. The PCC ramp-rate limit is set to $R_{grid} = 150$MW per 15-min interval, consistent with typical TSO requirements for large grid-connected resources.

IT power model parameters are calibrated using empirical measurements of NVIDIA A100 GPUs reported in [24]. Measured power–throughput operating points within the admissible throughput range $s \in [s_{\min}, 1]$, corresponding to a per-GPU electrical power range of approximately 150–250 W, are used to fit the quadratic surrogate model in (4). The calibration yields a peak throughput of $r_{peak} = 20{,}800$ workload units/s, a minimum admissible throughput $s_{\min} = 0.755$, and empirical coefficients $\alpha_0 = 1052.7$ , $\alpha_1 = -2288.6$ , and $\alpha_2 = 1469.4$ , which collectively capture the strictly convex dependence of electrical power consumption on normalized throughput within the calibrated operating region [25]. The benchmark case study represents a utility-scale AIDC with an aggregated IT power capacity of 250 MW. Based on the calibrated per-accelerator power–throughput relationship, this corresponds to an equivalent accelerator fleet size on the order of $10^6$units, used to scale the aggregate IT power and throughput to the system level. All computing throughput and electrical power quantities are linearly scaled from the calibrated per-GPU model to the system level. The efficiency of the internal power conditioning system is assumed to be $\eta_{IPCS}$ =0.95. In practice, the computational throughput and power characteristics of an AIDC may vary across workload types (e.g., training or inference). In this study, we focus on intra-day operation under a single homogeneous workload mode, and assume that all computing tasks within a given day follow the same calibrated power–throughput relationship (e.g., large-scale AI training) throughout the scheduling horizon.

The thermal behavior of the data hall is modeled using a first-order RC dynamics, with indoor temperature $T_{in}(t)$ constrained within [18,26]°C. The nominal energy input ratio is set to $\mathrm{EIR}_{nom}$=1/4.05 [26], with thermal capacitance $C_{th}$=120MWh /°C and thermal resistance $R_{th}$=0.2°C/MW. The maximum cooling extraction power is set to $Q_{cool}^{\max} = 250$MW, commensurate with the peak IT heat dissipation capacity. The on-site BESS is characterized by charging and discharging power limits $P_{ch}^{max} = P_{dis}^{max} = 400$MW, round-trip efficiencies $\eta_{ch} = \eta_{dis} = 0.95$, and state-of-charge bounds $E(t) \in [40,400]$ MWh. A cyclic SoC constraint is imposed with $E(T+1) = E(1)$ =200MWh. A battery cycling penalty $C_{\deg} = 30$AUD/MWh is applied to charging and discharging power to discourage excessive short-term cycling and ensure numerically smooth operation. Checkpoint availability is modeled using a repeating daily pattern $[0,0,0,1]$, indicating that shutdown actions are permitted once every hour. A large penalty $M_{\mathrm{RT}} = 10^6$ is assigned to the workload under-delivery slack to enforce the day-ahead committed workload as a hard priority in real-time operation.

The case study uses three synchronized time series at 15-minute resolution: electricity prices, ambient temperature, and system background demand. Electricity price and demand data are obtained from the Australian Energy Market Operator (AEMO) New South Wales (NSW) market for January 2026, while ambient temperature data are taken from historical weather observations in Sydney for the same period. The background demand profile is mapped onto the IEEE 39-bus transmission network through proportional scaling of nodal loads. Under "connect-and-manage" practices, power exchange limits are computed by the TSO in real time and communicated to the AIDC for execution. Although the IEEE 39-bus system is used in this paper for modeling purposes, in practice an AIDC does not have access to the actual transmission network topology or parameters. Therefore, it is infeasible for the AIDC to directly simulate power exchange limits, and it can only rely on historical power exchange limits issued by the TSO to learn the relationship between $z(t)$and the power exchange limits. In this paper, a conditional time-series generative adversarial network (CTSGAN) [27] is trained using a full year of data from 2025. Each training sample consists of a set of publicly observable system features $z(t)$ and the corresponding daily trajectories of upper and lower PCC power exchange limits. In the DA stage, the trained CTSGAN is used to predict the distribution of admissible PCC power-exchange limit trajectories for the upcoming operating day, conditional on the observed feature sequence $z(t)$. For each study day, the CTSGAN generates an initial ensemble of $N_{\mathrm{raw}} = 200$ admissible PCC limit trajectories. To control the level of conservatism in workload commitment, a coverage level of $1 - \alpha = 90\%$ is adopted. The generated trajectories are ranked using a trajectory-level risk metric reflecting the overall tightness of the admissible power limits, and the most extreme 10% are discarded. This results in a reduced scenario set $\Psi$ containing 180 representative trajectories.

The retained scenario set $\Psi$ is used exclusively in the day-ahead workload commitment problem, while the real-time stage operates on the realized PCC power-exchange limits computed and communicated by the TSO. For the real-time receding-horizon delivery assurance, the prediction horizon is set to H = 20 control intervals (5 hours). The proposed day-ahead and real-time formulations result in mixed-integer linear programs (MILPs). All operational constraints are linear after standard reformulations, and the convex power–throughput relationship in the IT power consumption model is approximated using a standard piecewise-linear representation. The resulting MILPs are efficiently solved using the commercial solver Gurobi.

The case studies are organized as a sequence of six cases, each addressing a specific aspect of BESS-assisted grid-aware AIDC operation. **Case 1** examines how BESS enables higher credible day-ahead workload commitment on a representative congested day. **Case 2** extends this analysis through a parametric study to reveal the regime-dependent marginal value of BESS for day-ahead commitment under different levels of transmission congestion. **Case 3** evaluates whether the committed workload can be reliably delivered in real time under realized PCC power exchange limits and price signals. **Case 4** further investigates how the operational role of BESS transitions from feasibility-oriented support to economy-driven flexibility as transmission constraints are relaxed. **Case 5** focuses on the impact of BESS energy capacity on IT–cooling coordination and physical continuity under severe grid constraints. Finally, **Case 6** examines how checkpoint frequency shapes the feasibility-oriented use of BESS in preserving AI workload continuity.

### *B. Case 1: BESS Enables Higher Credible Day-Ahead Workload Commitment under Transmission Congestion*

Since the objective of the DA stage is to maximize the committed workload $W_{DA}^{*}$, this subsection examines the impact of introducing a BESS on DA workload commitment. The case study is conducted using data from January 7, 2026. Fig. 2 illustrates a representative admissible PCC power-exchange limit trajectory selected from the day-ahead scenario set, which exhibits two pronounced congestion periods, namely time slots 56–69 and 76–92, indicating severe transmission constraints. A comparison of Fig. 2(a) and Fig. 2(b) shows that the AIDC without BESS support is only able to operate during a brief window around time slot 70, while remaining idle for most of the congested periods. With BESS support, however, the AIDC sustains an operating level of approximately 175 MW for most of the day, including time slots when the admissible PCC exchange range collapses to 0 MW. Representative examples include time slots 61–67 and 78–83, during which the no-BESS AIDC is completely shut down, whereas the BESS-assisted AIDC continues operating by relying on battery discharge.

The resulting throughput decisions are illustrated in Fig. 2(c) and Fig. 2(d). With BESS support, the normalized throughput $s(t)$ drops to zero at only four isolated time slots over the entire day. During other power-limit-constrained periods, the AIDC remains operational at an effective throughput factor of approximately $s = 0.83$, indicating throughput reduction rather than complete shutdown. In contrast, without BESS, the AIDC is unable to operate for a total of 31 time slots, corresponding to 7.75 hours of complete service interruption under the selected scenario.

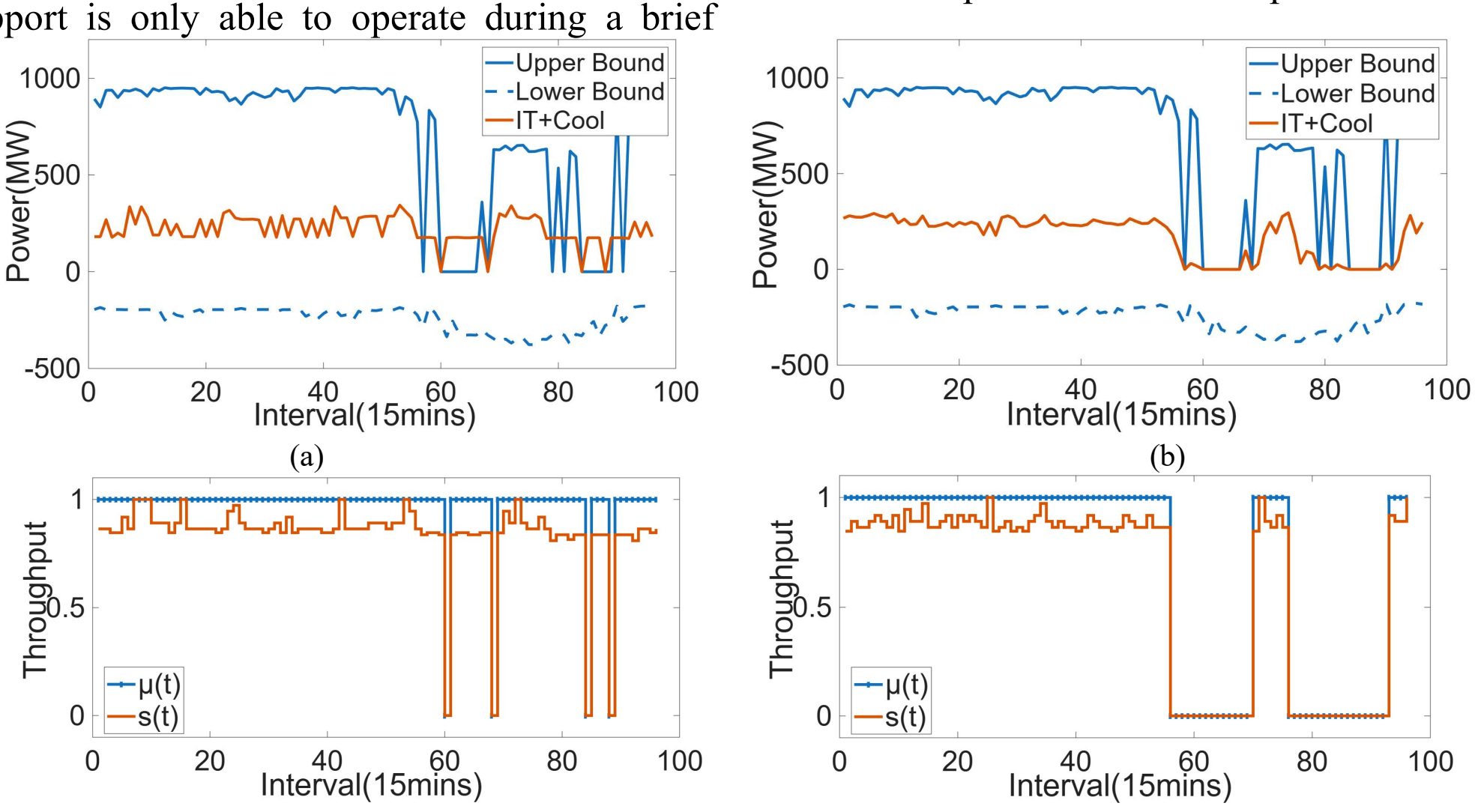

Fig. 2 AIDC–Grid Power Exchange and Throughput Decisions: (a) interaction with BESS, (b) interaction without BESS; (c) throughput decisions with BESS, (d) throughput decisions without BESS.

### *C. Case 2: Regime-Dependent Value of BESS for Day-Ahead Commitment*

To further investigate the impact of BESS on $W_{DA}^{*}$ under different network constraint levels, the line thermal limits of the IEEE 39-bus transmission test system are parametrically scaled from 1.0 to 1.5. In addition, different BESS configurations are considered, with energy capacities ranging from 200 MWh to 800 MWh, and the corresponding day-ahead $W_{DA}^{*}$ are computed. The results are summarized in Fig. 3.

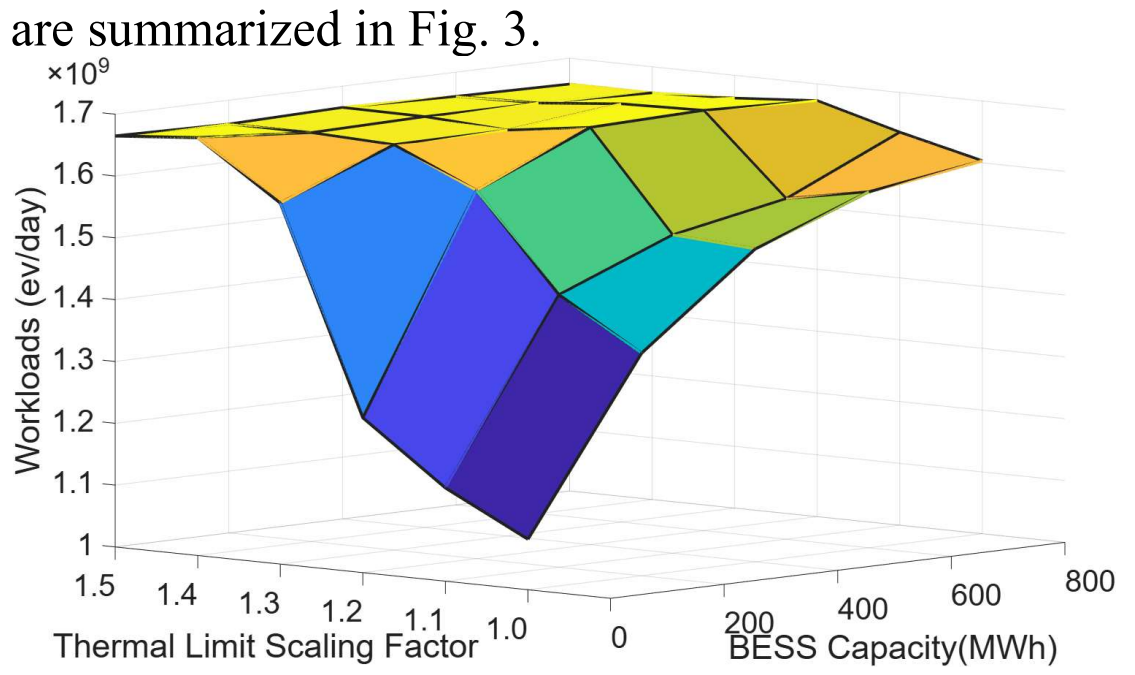

Fig. 3. Enhancement of $W_{DA}^{*}$ by BESS under Different Transmission Constraints

The results indicate that when transmission constraints are relatively tight (scaling factors of 1.0–1.2), the contribution of BESS to enhancing $W_{DA}^{*}$ is particularly pronounced. For example, at a scaling factor of 1.0, increasing the BESS capacity from 200 MWh to 800 MWh raises $W_{DA}^{*}$ from approximately $1.36 \times 10^{9}$ units to $1.60 \times 10^{9}$ units, demonstrating the effectiveness of BESS in alleviating commitment bottlenecks under transmission-limited conditions. By contrast, at scaling factors of 1.4 and 1.5, the values of $W_{DA}^{*}$ corresponding to different BESS capacities become highly similar, indicating a transition from a network-constrained regime to a computation-limited regime, where further increases in BESS size yield diminishing commitment benefits. These observations highlight that the value of BESS in the proposed framework is highly context-dependent: it plays a critical buffering role under network thermal congestion, while offering limited marginal benefit when transmission capacity is sufficiently relaxed.

### *D. Case 3: Real-Time Delivery Assurance of Day-Ahead Commitments*

The DA committed workload $W_{DA}^{*}$ is taken as an exogenous input to the RT optimization stage. All parameters and system settings are kept identical to those in the previous subsection. Fig. 4 (left axis) shows the evolution of the remaining workload over time. With BESS support, the AIDC operates under a significantly higher day-ahead workload commitment of $1.49 \times 10^{9}$, inherited from the DA stage, representing a 38.26% increase compared to the no-BESS case ($1.08 \times 10^9$), while still fully delivering the committed workload in the RT stage. This result confirms that the higher DA commitment enabled by BESS remains operationally feasible under realized interconnection limits and disturbances. Fig. 4 (right axis) further reports the total daily energy cost for the two cases. The BESS-assisted AIDC incurs an energy cost of $1.92 \times 10^5$ AUD, compared to $1.51 \times 10^5$ AUD in the no-BESS case. When normalized by the delivered workload, the BESS-assisted AIDC achieves an average computational efficiency of 7,798.1 workload units per AUD, whereas the no-BESS configuration achieves 7,142.0 units per AUD. The background shading in Fig. 4 depicts the normalized real-time electricity price profile, highlighting periods when exporting power is economically favorable. In particular, during time slots 76–81, the BESS-assisted AIDC injects power into the grid and yields a revenue of approximately $0.4987 \times 10^5$ AUD, which partially offsets the overall energy cost. In contrast, the no-BESS AIDC is forced to shut down during this period due to the combined effect of tight transmission thermal constraints and elevated electricity prices.

To further evaluate delivery robustness under uncertainty, a multi-day simulation is conducted over a 31-day horizon from January 1 to January 31, during which the DA–RT procedure is executed sequentially each day. Without BESS, workload under-delivery occurs on 17 days, whereas this number is reduced to only 5 days when BESS is available. Over the entire evaluation period, the average under-delivery with BESS is 91,597.6 units, corresponding to just $6.04 \times 10^{-5}$ of the total committed workload. By comparison, the no-BESS case exhibits a substantially larger average under-delivery of 508,022.7 units ($4.7 \times 10^{-4}$ of the total workload). In terms of worst-case performance, the maximum under-delivery is reduced from 2,133,445 units without BESS to 954,634 units with BESS. These aggregate statistics indicate that BESS substantially improves robustness against inter-day uncertainty and enhances the ability of the RT controller to maintain reliable workload delivery.

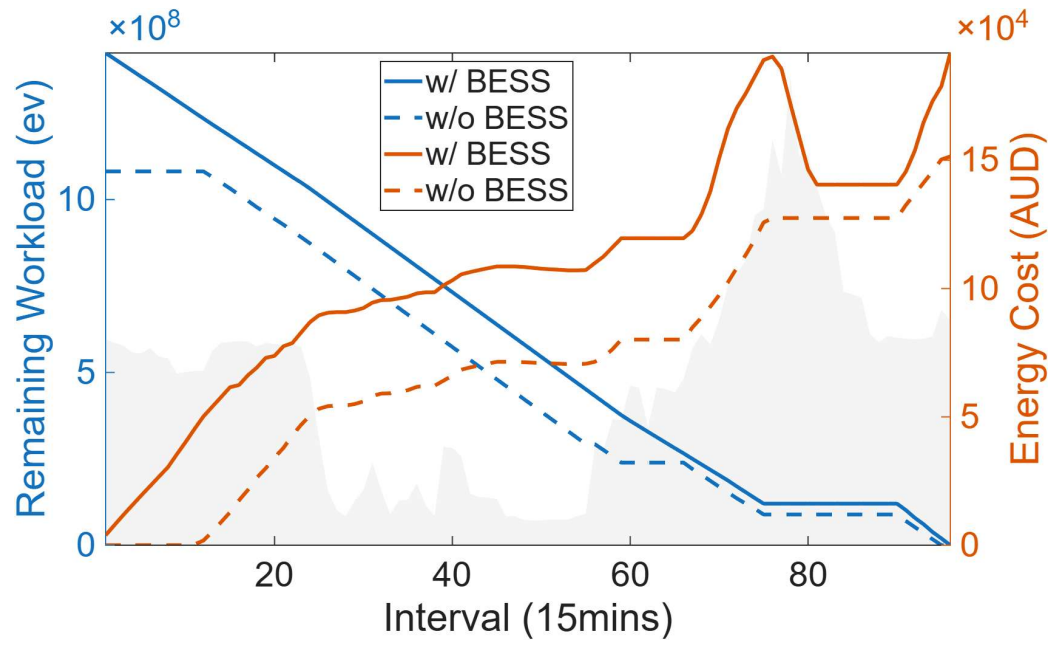

Fig. 4. RT Delivery Performance with and without BESS

### E. Case 4: Role Transition of BESS from Feasibility Support to Economic Flexibility

While the previous subsection focused on how transmission constraints shape the day-ahead commitment capacity, this subsection shifts attention to the real-time operational role of BESS under the same network conditions. In this analysis, transmission line thermal limits are progressively relaxed up to 1.5 times their nominal values, which in turn expands the feasible PCC power exchange region, while the BESS power and energy capacities are kept fixed.

Table I summarizes the impact of this relaxation on workload delivery and energy cost. It can be found that, as the thermal limit scaling factor increases from 1.0 to 1.4, transmission congestion is gradually alleviated but remains binding during critical operating periods. Under these conditions, the BESS primarily serves as a feasibility-oriented support resource, discharging during network-constrained time slots to compensate for insufficient admissible grid import and thereby sustaining real-time workload delivery. This operational role enables the delivered workload to increase from $1.51 \times 10^9$ to $1.64 \times 10^9$ units. However, because BESS dispatch in this regime is driven mainly by feasibility requirements rather than price signals, the associated energy cost rises substantially.

TABLE I
IMPACT OF TRANSMISSION THERMAL LIMITS ON AIDC

| Thermal Limit Scaling Factor | Workload Delivery ($10^9$ ev/day) | Energy Cost ($10^5$ AUD/day) |
|---|---|---|
| 1.0 | 1.51 | 1.92 |
| 1.1 | 1.53 | 2.16 |
| 1.2 | 1.60 | 2.44 |
| 1.3 | 1.64 | 3.23 |
| 1.4 | 1.66 | 3.62 |
| 1.5 | 1.66 | 3.54 |

When the thermal limits are further relaxed to a scaling factor of 1.5, the resulting PCC power exchange limits no longer bind the AIDC operation. Consequently, the role of the BESS shifts from feasibility-driven support to economy-driven flexibility provision. As illustrated in Fig. 5, unlike the case with a scaling factor of 1.0 where the BESS is required to discharge during critical time slots (60–68) to support workload delivery, the BESS under a scaling factor of 1.5 preserves this energy and reallocates it to periods with higher electricity prices, notably around time slot 80. This shift in operating behavior allows the BESS to exploit price-driven dispatch opportunities, leading to a reduction in overall energy cost without further increasing workload delivery. These results highlight that transmission constraint relaxation fundamentally alters the role of BESS in AIDC operation, from ensuring feasibility under network congestion to enhancing economic efficiency once grid constraints are non-binding.

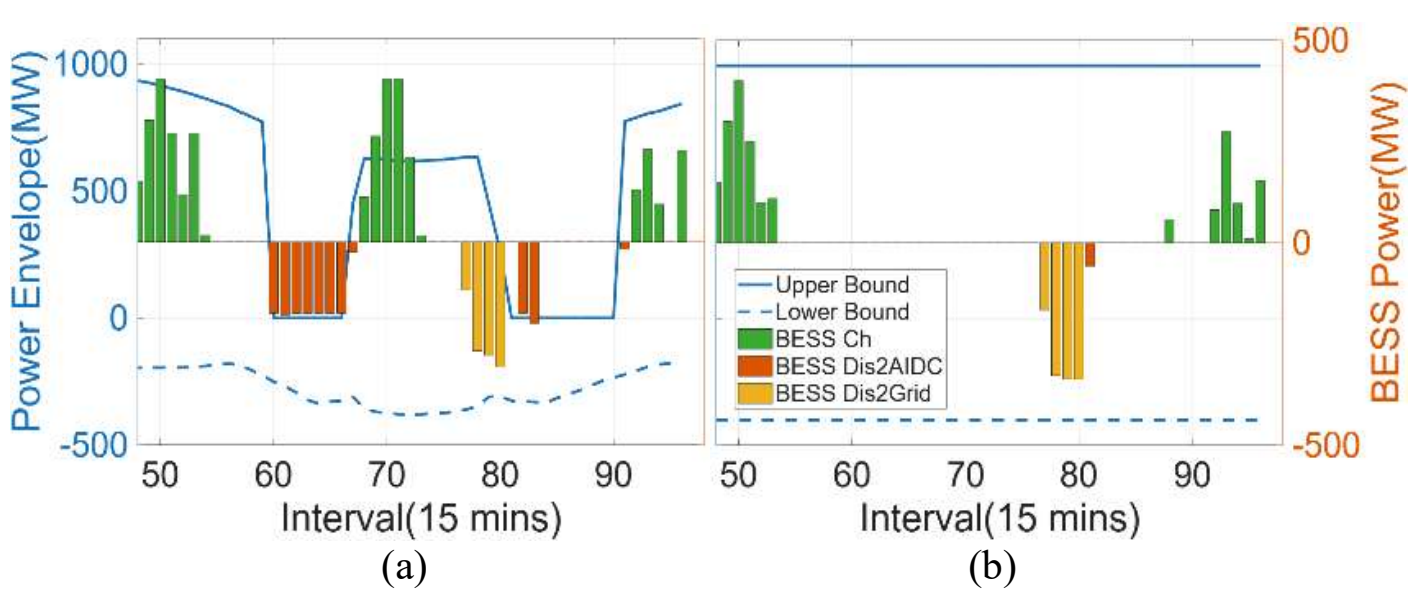

Fig. 5 BESS Behavior under Different Thermal Limit Scaling Factors: (a) 1.0, and (b) 1.5.

### F. Case 5: BESS Energy Capacity as a Key Enabler of IT–Cooling Coordination

To investigate how the energy capacity of the BESS affects the coordination between the IT and cooling subsystems, we conduct a sensitivity analysis by scaling the baseline BESS energy capacity by ±20%, while keeping the maximum charging and discharging power unchanged. The resulting system behaviors are illustrated in Fig. 6, where Fig. 6(a) illustrates the IT power consumption together with the BESS power supporting the AIDC, and Fig. 6(b) shows the temperature trajectory and the corresponding cooling power.

As shown in Fig. 6(a), when the BESS energy capacity is reduced to 320 MWh, the storage system is unable to effectively support the AIDC during the grid-constrained period around time slot 60. As a result, the IT power abruptly drops from approximately 172 MW to zero, indicating a complete shutdown of the computing cluster. During this period, the BESS no longer supplies power to the AIDC, as the limited stored energy is insufficient to sustain operation under binding grid constraints. By contrast, when the BESS energy capacity is increased to 480 MWh, the AIDC is able to maintain the IT cluster in operation at a reduced level during the same congestion period. In this case, the BESS continuously discharges at approximately 175 MW to support the AIDC, with the supplied power being almost entirely allocated to IT computation. This comparison highlights that sufficient BESS energy capacity is critical for preserving computing continuity when grid power exchange is severely restricted.

Fig. 6(b) shows that the cooling power trajectories exhibit broadly similar trends across different BESS capacities, and no substantial divergence is observed in the overall cooling profiles. Correspondingly, the indoor temperature trajectories remain largely aligned, except for a noticeable temperature shift occurring around time slot 61. By jointly examining Fig. 6(a) and Fig. 6(b), it can be seen that this temperature deviation is induced by the change in IT operation rather than by active differences in cooling control. A particularly notable observation is that, prior to the onset of grid congestion (before time slot 60), the cooling system operates close to its maximum power, reducing the indoor temperature from 26 °C to approximately 18 °C. During the subsequent congestion period, the limited electrical power, supplied primarily by BESS discharging, is almost entirely prioritized for IT computation, while the cooling system is largely suspended, except for brief activations around time slots 73 and 76. Throughout this phase, the indoor temperature rises gradually but remains within the admissible range, after which normal cooling operation resumes once grid constraints are relieved. This behavior illustrates an implicit pre-cooling strategy enabled by sufficient BESS energy capacity, whereby thermal inertia is exploited to reallocate electrical power from cooling to IT computation during periods of severe grid congestion.

Fig. 6 Effect of BESS Energy Capacity on AIDC Operation. (a) IT Power and BESS Support, and (b) Indoor Temperature and Cooling Power

## *G. Case 6: Checkpoint Frequency Shapes the Feasibility-Oriented Use of BESS*

Fig. 7 compares the real-time operation of the AIDC under three periodic checkpointing schemes, with checkpoint intervals of 30 minutes, 1 hour, and 2 hours, respectively. The BESS capacity and grid-interconnection conditions are kept identical, and only the checkpoint availability pattern is varied. In Fig. 7(a), where checkpoint opportunities occur every 30 minutes, the computing cluster exhibits high operational flexibility. Under this dense checkpointing regime, the AIDC can opportunistically shut down at suitable time instants without requiring sustained battery support. As a result, the BESS is rarely utilized for feasibility-driven support, and IT operation is primarily coordinated through timely shutdown actions aligned with checkpoint availability. In contrast, Fig. 7(b) corresponds to a checkpoint interval of 1 hour. We can find that the shutdown opportunity occurs at time slot 76. During the preceding time slot 73–75, the AIDC is unable to shut down due to checkpointing constraints, while grid-interconnection limits become binding. To bridge this gap, the BESS provides substantial discharging support in the range of 181–189 MW, enabling the IT cluster to remain operational until the next checkpoint is reached, after which a controlled shutdown is executed. A similar but more pronounced behavior is observed in Fig. 7(c), where the checkpoint interval is extended to 2 hours. Here, the AIDC must remain active over a longer duration before the next admissible shutdown point at time slot 80. Accordingly, the BESS supplies sustained support over four consecutive time steps, effectively compensating for insufficient grid power and preserving workload continuity throughout the extended checkpoint-constrained period.

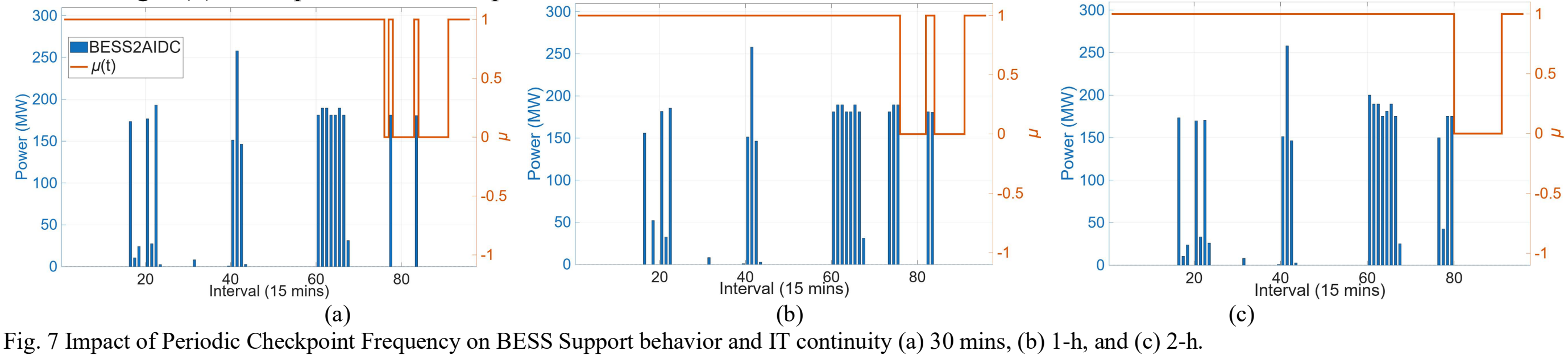

Fig. 7 Impact of Periodic Checkpoint Frequency on BESS Support behavior and IT continuity (a) 30 mins, (b) 1-h, and (c) 2-h.

These results demonstrate that the utilization pattern of the BESS is strongly shaped by checkpoint frequency. When checkpointing is frequent, workload-level flexibility alone is sufficient to handle grid constraints. As checkpoint opportunities become sparser, however, the BESS increasingly serves as a continuity-bridging resource, providing feasibility-oriented support that allows the AIDC to span grid-constrained intervals until checkpoint-compliant shutdown becomes possible.

## VI. CONCLUSION

This paper presents a battery-assisted operational framework for hyperscale AIDCs operating under "connect-and-manage" practices. By explicitly positioning the on-site BESS as a physical interface between fast internal computing – thermal dynamics and externally imposed power exchange limits at the point of common coupling, the proposed framework addresses the fundamental mismatch between internal operational flexibility and grid security requirements. Comprehensive case studies demonstrate that BESS substantially enhances both day-ahead workload commitment capability and real-time delivery reliability under transmission congestion. The results further reveal a regime-dependent role transition of BESS: when transmission constraints are binding, BESS primarily acts as a feasibility-oriented continuity-bridging resource to sustain workload execution, whereas under relaxed network conditions it shifts to an economy-driven flexibility provider that exploits price signals to improve operational efficiency. Moreover, the interaction among BESS energy capacity, thermal inertia, and checkpoint-constrained computing is shown to be critical for preserving AI workload continuity under grid-constrained operation.

Overall, the proposed framework offers a physically grounded and operationally tractable approach for integrating large-scale AIDCs into constrained transmission networks. The findings provide actionable insights for the co-design of future power systems and computation-intensive infrastructures, particularly in regulatory and operational contexts characterized by connect-and-manage interconnection practices.

## APPENDIX A: ILLUSTRATIVE DC POWER FLOW RELATIONS FOR POWER EXCHANGE LIMITS DERIVATION

This appendix provides an illustrative example of how transmission thermal constraints may lead to admissible power exchange limits at the PCC. Other operational constraints, such as voltage limits, or stability margins can be accommodated within the same limit-based abstraction. The active power flow on the transmission line $(k, j) \in \mathcal{L}$, denoted as $P_{line,kj}(t)$, is determined by the phase angle difference and the line susceptance $B_{kj}$:

$$P_{line,kj}(t) = B_{kj}\left(\theta_k(t) - \theta_j(t)\right) \quad \text{(A1)}$$

where $\theta_k(t)$ is the voltage phase angle at bus $k$. The reference bus phase angle is fixed to zero: $\theta_{ref}(t) = 0$. To ensure operational safety, the power flow on any transmission line must not exceed its thermal stability limit $F_{max,kj}$:

$$-F_{max,kj} \le P_{line,kj}(t) \le F_{max,kj} \quad \text{(A2)}$$

Based on these relations, the TSO evaluates background power flows and congestion conditions under forecast operating scenarios, and determines the admissible range of net active power exchange at the PCC that does not violate line thermal limits or other security constraints.

## REFERENCES

[1] J. Mills, "NVIDIA Launches Omniverse DSX Blueprint, Enabling Global AI Infrastructure Ecosystem to Build Gigawatt-Scale AI Factories," *NVIDIA Blog*, 2025.

[2] X. Chen, X. Wang, A. Colacelli, M. Lee, and L. Xie, "Electricity demand and grid impacts of AI data centers: Challenges and prospects," *arXiv preprint arXiv:2509.07218*, 2025.

[3] I. E. Agency, AI is set to drive surging electricity demand from data centres while offering the potential to transform how the energy sector works, *IEA News Release*, 2025.

[4] L. Xie, N. Li, and H. V. Poor, "Sustainable electrification in the era of AI," *Nature Reviews Electrical Engineering,* vol. 1, no. 8, pp. 493-494, 2024.

[5] PJM, "PJM Board Outlines Plans To Integrate Large Loads Reliably," *PJM Inside Lines*, 2026.

[6] J. Jian, J. Zhao, H. Ji, L. Bai, J. Xu, P. Li, J. Wu, and C. Wang, "Supply restoration of data centers in flexible distribution networks with spatial-temporal regulation," *IEEE Transactions on Smart Grid,* vol. 15, no. 1, pp. 340-354, 2023.

[7] X. Yin, C. Ye, Y. Ding, and Y. Song, "Exploiting internet data centers as energy prosumers in integrated electricity-heat system," *IEEE Transactions on Smart Grid,* vol. 14, no. 1, pp. 167-182, 2022.

[8] T. Wan, Y. Tao, J. Qiu, and S. Lai, "Internet data centers participating in electricity network transition considering carbon-oriented demand response," *Applied Energy,* vol. 329, pp. 120305, 2023.

[9] T. Jin, L. Bai, M. Yan, and X. Chen, "Unlocking Spatio-Temporal Flexibility of Data Centers in Multiple Regional Peer-to-Peer Energy Transaction Markets," *IEEE Transactions on Power Systems*, 2025.

[10] H. Wang, J. Huang, X. Lin, and H. Mohsenian-Rad, "Exploring smart grid and data center interactions for electric power load balancing," *ACM SIGMETRICS Performance Evaluation Review,* vol. 41, no. 3, pp. 89-94, 2014.

[11] B. Wan, M. Han, Y. Sheng, Y. Peng, H. Lin, M. Zhang, Z. Lai, M. Yu, J. Zhang, and Z. Song, "{ByteCheckpoint}: A Unified Checkpointing System for Large Foundation Model Development." pp. 559-578.

[12] S. Qi, L. Niu, and Z. Wu, "Study on AI Data Center Infrastructure Sustainable Deployment and Standardization." pp. 1480-1487.

[13] S. Karimi, P. Musilek, and A. M. Knight, "Dynamic thermal rating of transmission lines: A review," *Renewable and Sustainable Energy Reviews,* vol. 91, pp. 600-612, 2018.

[14] Z.-P. Yuan, P. Li, Z.-L. Li, and J. Xia, "Data-driven risk-adjusted robust energy management for microgrids integrating demand response aggregator and renewable energies," *IEEE Transactions on Smart Grid,* vol. 14, no. 1, pp. 365-377, 2022.

[15] S. An, J. Qiu, J. Lin, Z. Yao, Q. Liang, and X. Lu, "Planning of a multi-agent mobile robot-based adaptive charging network for enhancing power system resilience under extreme conditions," *Applied Energy,* vol. 395, pp. 126252, 2025.

[16] R. R. Ahrabi, A. Mousavi, E. Mohammadi, R. Wu, and A. K. Chen, "AI-Driven Data Center Energy Profile, Power Quality, Sustainable Sitting, and Energy Management: A Comprehensive Survey."

[17] X. Lu, J. Zhao, J. Qiu, C. Zhang, G. Lei, and J. Zhu, "Deposit and withdraw: Reinforcement learning-based incentive design for shared energy storage," *Energy Conversion and Economics,* vol. 6, no. 5, pp. 308-323, 2025.

[18] T. Wan, J. Qiu, Y. Tao, S. Lai, and R. Mao, "Flexible Energy Storage System and Renewable Energy Planning for Sustainable Internet Data Center Considering Temporal and Spatial Load Regulation," *IEEE Transactions on Industry Applications*, 2025.

[19] Y. Zhang, B. Zou, X. Jin, Y. Luo, M. Song, Y. Ye, Q. Hu, Q. Chen, and A. C. Zambroni, "Mitigating power grid impact from proactive data center workload shifts: A coordinated scheduling strategy integrating synergistic traffic-data-power networks," *Applied Energy,* vol. 377, pp. 124697, 2025.

[20] H. Yuan, J. Bi, S. Li, J. Zhang, and M. Zhou, "An improved LSTM-based prediction approach for resources and workload in large-scale data centers," *IEEE Internet of Things Journal,* vol. 11, no. 12, pp. 22816-22829, 2024.

[21] S. Liu, T. Zhao, X. Liu, Y. Li, and P. Wang, "Proactive resilient day-ahead unit commitment with cloud computing data centers," *IEEE Transactions on Industry Applications,* vol. 58, no. 2, pp. 1675-1684, 2022.

[22] S. Hong, and H. Kim, "An integrated GPU power and performance model." pp. 280-289.

[23] U. S. D. o. Energy, *Technical Support Document: Energy Efficiency Program for Consumer Products – ASHRAE Standard 90.1-2010 Final Rule, Chapter 4: Energy Use Characterization.*, 2012.

[24] S. B. Vanestan, Optimizing the operation of GPUs to reduce power consumption, *Sharif University of Technology (SUT)*, 2025.

[25] J. You, J.-W. Chung, and M. Chowdhury, "Zeus: Understanding and optimizing GPU energy consumption of DNN training." pp. 119-139.

[26] M. Borkowski, and A. K. Piłat, "Customized data center cooling system operating at significant outdoor temperature fluctuations," *Applied Energy,* vol. 306, pp. 117975, 2022.

[27] X. Lu, J. Qiu, G. Lei, and J. Zhu, "An interval prediction method for day-ahead electricity price in wholesale market considering weather factors," *IEEE Transactions on Power Systems,* vol. 39, no. 2, pp. 2558-2569, 2023.